\documentclass[preprint]{elsarticle}
\usepackage{graphicx}
\usepackage{amssymb,amsmath}
\usepackage{subfig}
\usepackage{color}
\usepackage{hyperref}
\usepackage{xspace}

\newcommand{\nc}[1]{\newcommand #1}
\newcommand{\rnc}[1]{\renewcommand #1}
\nc{\myquote}[1]{`#1'}
\nc{\x}[1]{\mbox{#1}}
\rnc{\matrix}[2]{\left[\!\!\begin{array}{#1} #2\end{array}\!\!\right]}
\rnc{\vector}[1]{\matrix{c}{#1}}
\nc{\suml}[2]{\sum \limits_{#1}^{#2}}
\nc{\real}[1]{\Re\lbrace #1 \rbrace}
\nc{\imag}[1]{\Im\lbrace #1 \rbrace}
\nc{\etal}{et~al.\,}
\nc{\cc}{\mathrm{c.c.}}
\nc{\g}[1]{\x{$#1$}}
\nc{\e}[2]{\begin{equation} #1 \label {eq:#2} \end{equation}}
\nc{\eal}[2]{\begin{equation} \begin{aligned} #1 \label {eq:#2} \end{aligned} \end{equation}}
\nc{\ea}[2]{\begin{eqnarray}
#1 \label {eq:#2}
\end{eqnarray}}
\nc{\inv}{^{-1}}
\nc{\tra}{^{\mathrm T}}
\nc{\herm}{^{\mathrm H}}
\nc{\fabstand}{\,}
\nc{\fp}{\fabstand.}
\nc{\fk}{\fabstand,}
\nc{\mm}[1]{\mathbf{#1}}
\nc{\mms}[1]{\boldsymbol{#1}}
\nc{\ie}{i.\,e.\xspace}
\nc{\eg}{e.\,g.\xspace}
\nc{\cf}{cf.\xspace}
\nc{\dd}{{\mathrm{d}}}
\nc{\ii}{{\mathrm{i}}}
\nc{\jj}{\ii}
\nc{\ee}{{\mathrm{e}}}
\nc{\fe}{f_{\mathrm{exc}}}
\nc{\fj}{f_{\mathrm{j}}}
\nc{\fr}{f_{\mathrm{c}}}
\nc{\ut}{w_{\mathrm{c}}}
\nc{\fnp}{N}
\nc{\kt}{k_{\mathrm{c}}}
\nc{\ktweak}{k_{\mathrm{c}^{\mathrm{weak}}}}
\nc{\ktstrong}{k_{\mathrm{c}^{\mathrm{strong}}}}
\nc{\pex}{p_{\mathrm{exc}}}
\nc{\ewonel}{\mm e_{w_1(\ell)}}
\nc{\frv}{\mm f_{\mathrm c}}
\nc{\sigrms}{\hat\sigma}
\nc{\sigcrit}{\hat\sigma_{\mathrm{crit}}}
\nc{\sigcriti}{\hat\sigma_{\mathrm{crit},i}}
\nc{\sigcritone}{\hat\sigma_{\mathrm{crit},1}}
\nc{\sigcrittwo}{\hat\sigma_{\mathrm{crit},2}}
\nc{\performance}{\hat\sigma_{\mathrm{crit,max}}}
\nc{\performanceref}{\hat\sigma_{\mathrm{crit,max},\mathrm{detuned}}}
\nc{\alphad}{\alpha_{\mathrm{d}}}
\nc{\betad}{\beta_{\mathrm{d}}}
\nc{\Deltaopt}{\Delta_{\mathrm{best}}}
\nc{\Deltaworst}{\Delta_{\mathrm{worst}}}
\nc{\wdone}{W_{\mathrm{diss},1}}
\nc{\wdthree}{W_{\mathrm{c},3}}
\nc{\fconone}{\hat f_{\mathrm{c},1}}
\nc{\fconthree}{\hat f_{\mathrm{c},3}}
\nc{\deffone}{d_{\mathrm{eff},1}}
\nc{\diag}{\operatorname{diag}}
\nc{\fref}[1]{\x{Fig.~\ref{fig:#1}}}
\nc{\frefs}[1]{\x{Figs.~\ref{fig:#1}}}
\nc{\frefo}[1]{\x{\ref{fig:#1}}}
\nc{\eref}[1]{\x{Eq.~(\ref{eq:#1})}}
\nc{\erefs}[1]{\x{Eqs.~(\ref{eq:#1})}}
\nc{\erefo}[1]{(\ref{eq:#1})}
\nc{\tref}[1]{\x{Tab.~\ref{tab:#1}}}
\nc{\sref}[1]{\x{Sec.~\ref{sec:#1}}}
\nc{\srefo}[1]{\ref{sec:#1}}
\nc{\srefs}[1]{\x{Sec.~\ref{sec:#1}}}
\nc{\ssref}[1]{\x{Subsec.~\ref{sec:#1}}}
\nc{\aref}[1]{\x{Appendix~\ref{asec:#1}}}
\nc{\fig}[3][tbh]{
\begin{figure}[#1]
\centering
\includegraphics{figs/#2}
\caption{#3\label{fig:#2}}
\end{figure}}
\nc{\figw}[3][tbh]{
\begin{figure*}[#1]
\centering
\includegraphics[width=1.0\textwidth]{figs/#2}
\caption{#3\label{fig:#2}}
\end{figure*}}
\nc{\figs}[4][tbh]{
\begin{figure*}[#1]
\centering
\includegraphics[scale=#4]{figs/#2}
\caption{#3\label{fig:#2}}
\end{figure*}}

\begin{document}

\begin{frontmatter}

\title{On the Efficacy of Friction Damping in the Presence of Nonlinear Modal Interactions}


\author[ila]{Malte Krack\corref{cor1}}
\ead{krack@ila.uni-stuttgart.de}
\address[ila]{Institute of Aircraft Propulsion Systems,
University of Stuttgart, 70569 Stuttgart, Germany, Tel.: +49-711-685-69391}

\author[affilbergman]{Lawrence A. Bergman}
\ead{lbergman@illinois.edu}
\address[affilbergman]{Department of Aerospace Engineering, University of Illinois at Urbana-Champaign, 104 S. Wright Street, Urbana, IL 61801, USA}

\author[affilvakakis]{Alexander F. Vakakis}
\ead{avakakis@illinois.edu}
\address[affilvakakis]{Department of Mechanical Science and Engineering, University of Illinois at Urbana-Champaign, 1206 W. Green Street, Urbana, IL 61801, USA}

\cortext[cor1]{Corresponding author}

\begin{abstract} 
This work addresses friction-induced modal interactions in jointed structures, and their effects on the passive mitigation of vibrations by means of friction damping. Under the condition of (nearly) commensurable natural frequencies, the nonlinear character of friction can cause so-called nonlinear modal interactions. If harmonic forcing near the natural frequency of a specific mode is applied, for instance, another mode may be excited due to nonlinear energy transfer and thus contribute considerably to the vibration response. We investigate how this phenomenon affects the performance of friction damping. To this end, we study the steady-state, periodic forced vibrations of a system of two beams connected via a local mechanical friction joint. The system can be tuned to continuously adjust the ratio between the first two natural frequencies in the range around the $1:3$ internal resonance, in order to trigger or suppress the emergence of modal interactions. Due to the re-distribution of the vibration energy, the vibration level can in fact be reduced in certain situations. However, in other situations, the multi-harmonic character of the vibration has detrimental effects on the effective damping provided by the friction joint. The resulting response level can be significantly larger than in the absence of modal interactions. Moreover, it is shown that the vibration behavior is highly sensitive in the neighborhood of internal resonances. It is thus concluded that the condition of internal resonance should be avoided in the design of friction-damped systems.
\end{abstract}

\begin{keyword}
friction damping \sep modal interactions \sep internal resonance \sep jointed structures 
\end{keyword}

\end{frontmatter}

\section{Introduction\label{sec:intro}}
Friction damping is a well-known means of achieving vibration reduction. The damping effect is due to the dissipative character of dry friction occurring in mechanical joints. For this purpose, friction interfaces may be either newly introduced to the structure or may already exist \eg in the form of bolted or riveted joints. Friction damping is particularly suited for the passive vibration reduction of lightly damped flexible structures. Various applications can be found in the field of aerospace structures, combustion engines or turbomachinery blades \cite{ferr1995,popp2003a}.\\
Dry friction is a nonlinear phenomenon: Depending on the vibration level of a jointed structure, the behavior in the local contact interfaces may range from sticking via micro-slip to macro-sliding. Hence, dry friction may induce the common nonlinear phenomena. In the context of steady-state harmonically-forced vibrations, these phenomena include: (a) The dependence of the resonance frequency and the effective damping on the vibration level, (b) the loss of stability of the fundamental periodic response giving rise to sub-harmonic, quasi-periodic or chaotic vibrations, and (c) the occurrence of so-called nonlinear modal interactions.\\
Nonlinear modal interactions refer to energy exchanges between two or more of a structure's modes of vibration caused by nonlinear effects. Thus, a mode that is not directly forced by an external source, may be excited indirectly by the action of nonlinear forces, and therefore contribute significantly to the overall vibration response. This occurs when the natural frequency of the indirectly forced mode is in a rational relation with the natural frequency of a directly forced mode \cite{nayf2000}, \ie, when the two modes are in \textit{internal resonance}. Under this condition, a suitable nonlinearity can initiate \textit{mode mixing} caused by energy transfers between the modes, even if their frequencies are well separated. Note that the necessary condition of internal resonance does not have to hold for the natural frequencies of the linearized system, but this condition can be satisfied for larger vibration levels of the nonlinear system when the natural frequencies change accordingly.\\
In the context of friction damping, numerical and experimental evidence of nonlinear modal interactions was reported, \eg, in \cite{whit1997,bert1998,chen2000,chen2001a,krac2013a}. Specifically, an important effect on the steady-state forced vibration response in the excitation frequency range around a particular natural frequency is commonly reported: The response of the indirectly forced mode assumes a local maximum. At the same time, the expected pronounced peak in the amplitude-frequency curve of the directly forced mode transforms to a double peak with a local minimum in between two local maxima. This phenomenon appears to be more profound if the joint's normal load fluctuates and the joint undergoes lift off/impacts as a consequence of vibrations \cite{chen2000,chen2001a,krac2013a}. Ferri \etal explored the energy transfers from low to high frequencies of friction damped systems under the condition of internal resonances, both in the steady-state forced response and the free vibrations \cite{Ferri.1992,Do.2005,Do.2008}.\\
Although nonlinear modal interactions are known to occur in friction-damped systems, their influence on the vibration reduction performance is unknown and yet has to be studied. It should be noted that nonlinear modal interactions are known to play the essential role in the vibration control concept of targeted energy transfer \cite{vaka2008b}. The key of this concept is that the vibration energy is passively and irreversibly transferred from a primary mode to either a strongly nonlinear attachment or a higher-frequency modes, where it is dissipated more efficiently. 
The purpose of the present study is thus to investigate how nonlinear modal interactions affect the efficacy of friction damping. To this end, we study a system composed of two cantilevered beams connected via a friction joint, as presented in \sref{model}. The system can be tuned so that the beams' first bending modes are in $1:3$ internal resonance. It is demonstrated in \sref{modal_interactions} how and under what conditions the friction nonlinearity induces the intended nonlinear modal interactions. The consequences for the vibration reduction performance are quantified and explained in \sref{detuning}. This article ends with concluding remarks in \sref{conclusions}.

\section{Model description\label{sec:model}}
\subsection{Problem formulation}
\figw[t!]{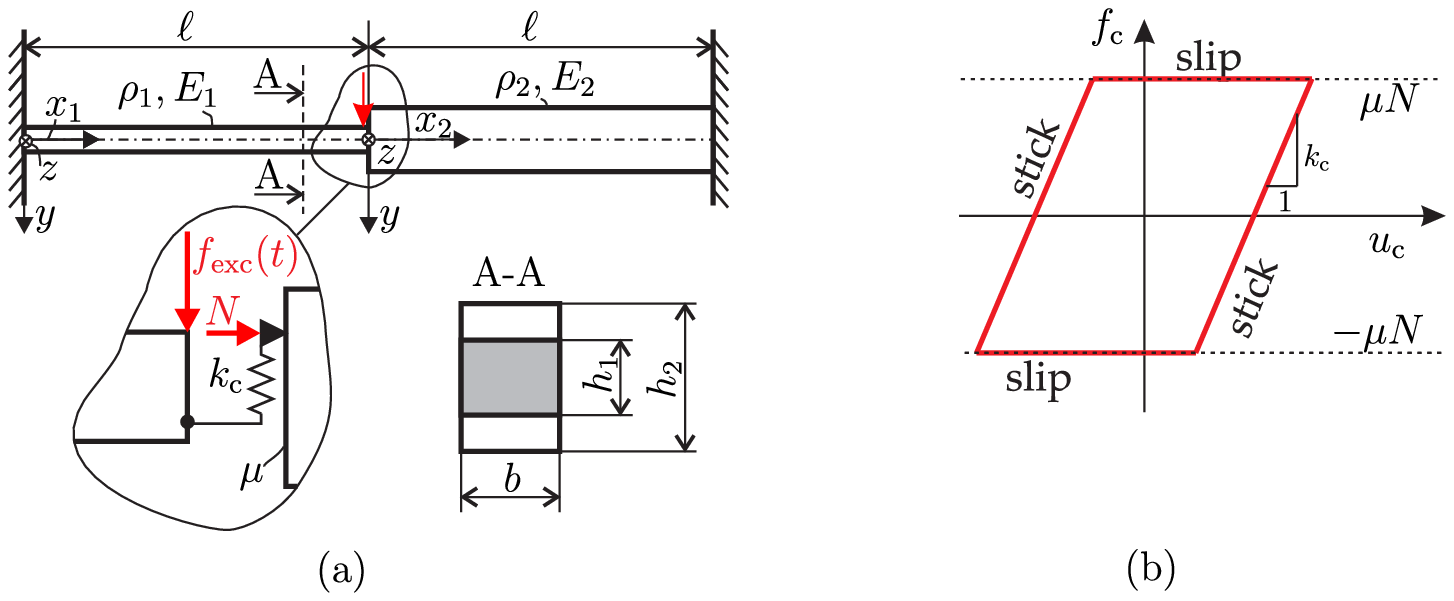}{Investigated model: (a) two beams connected via friction joint, (b) elastic Coulomb friction model; $E_1=6\cdot 10^6=E_2$, $\rho_1=2500=\rho_2$, $l_1=1.0=l_2$, $b_1=0.02=b_2$, $h_1=0.02$}
Consider the system of two beams connected via a friction joint, depicted in \fref{figure01}a. Bending vibrations with the deflection $w(x,t)$ in $y$-direction are considered. For convenience, the problem is divided into two sections, one for each beam, with local coordinates $x_1$ and $x_2$, respectively. In accordance with the classical Euler-Bernoulli beam theory\footnote{small deflections, plane sections remain plane, constant length neutral axis, linear-elastic material}, the equations of motion of the system can be written as,
\ea{
  & \left(E_iI_i w_i^{\prime\prime}(x_i,t)\right)^{\prime\prime} + \rho_i A_i \ddot w_i(x_i,t) = 0\fk\,\, i=1,2\fk\,\, 0<x_i<\ell \label{eq:local_equilibrium}\\
  & w_1(0,t) = 0 = w_1^\prime(0,t)\fk\,\, w_2(\ell,t)=0=w_2^\prime(\ell,t)\label{eq:clamped_ends}\\
  & \left(E_1I_1 w_1^{\prime\prime}\left(\ell,t\right)\right)^{\prime}+\fe(t) = \fj = -\left(E_2I_2 w_2^{\prime\prime}\left(0,t\right)\right)^{\prime}\fk \,\, w_1^{\prime\prime}(\ell,t) = 0 = w_2^{\prime\prime}(0,t)\label{eq:force_equilibrium}\\
  & \fj = \fr\left[\left(w_1(\ell)-w_2(0)\right)\right]\label{eq:joint_mechanics}\\
  & \fr\left[\ut\right]:\,\, \dd\fr = \begin{cases} \kt\dd \ut & \left|\fr+\kt\dd \ut\right|\leq\mu\fnp\\ 0 & \text{otherwise}\end{cases} \fk \label{eq:elastic_coulomb}
}{dont_reference_me_you_idiot}
where \eref{force_equilibrium} denotes the force balance at the joint coupling the two beams, and \erefs{joint_mechanics}-\erefo{elastic_coulomb} provide the constitutive force-deformation law for the frictional connection. Note that local spatial coordinates are used for each of the two beams.
In \erefs{local_equilibrium}-\erefo{elastic_coulomb}, prime denotes differentiation with respect to local coordinate $x_1$ or $x_2$, and overdot denotes differentiation with respect to time, and $A_i = b_ih_i$ and $I_i = \frac{b_ih_i^3}{12}$ are the area and the moment of inertia of the rectangular cross section, respectively. Note that the elastic Coulomb friction law is considered at the frictional connection (governed by \eref{elastic_coulomb}). The elastic Coulomb law relates the friction force $\fr$ and the relative displacement $\ut$. This relationship depends on the joint stiffness $\kt$ and the limiting friction force $\mu\fnp$, which is the product of the friction coefficient $0\leq \mu\leq 1$ and the joint normal force $\fnp>0$. The hysteretic dependence of the friction force on the relative joint deformation $\ut$ is expressed by the operator notation $\left[\ut\right]$. An example for a simple hysteresis cycle is depicted in \fref{figure01}b. The joint normal force is well known to have an essential influence on the friction damping performance. Therefore, its value was varied during this work, as detailed later.

\subsection{Spatial discretization}
System \erefo{local_equilibrium}-\erefo{elastic_coulomb} is spatially discretized by a finite element procedure. Each beam section is equidistantly divided into $N_{\mathrm{e},1}=N_{\mathrm{e},2}=20$ elements. Standard beam elements with consistent element matrices were utilized with two nodes per element and two degrees of freedom per node, one for the deflection and one for the inclination. This results in a set of second-order ordinary differential equations governing the vector of generalized coordinates $\mm u(t)$,
\e{\mm M\ddot{\mm u}(t) + \mm C_{\mathrm{mod}}\dot{\mm u}(t) + \mm K\mm u(t) + \frv\left[\mm u(t)\right] = \pex\ewonel\cos\Omega t\fp}{eqm} 
The total number of generalized coordinates is $N=2\left(N_{\mathrm{e},1}+N_{\mathrm{e},2}\right)=40$. A harmonic external forcing of magnitude $\pex$ and frequency $\Omega$ is considered in this study. $\ewonel$ is the unit vector with all components zero except the one associated with the displacement $w_1(\ell)$ at location $x_1=\ell$. A linear viscous damping term was introduced in \eref{eqm} to take into account the material damping effect. It is defined as modal damping,
\e{\mms\Phi\tra \mm C_{\mathrm{mod}} \mms\Phi = \diag\left(2D_i\omega_i\right)\fk}{modal_damping}
in the modal space associated to the linear system without joint. The quantities $\omega_i$ and $D_i$ are the $i$-th modal frequency and damping ratio, respectively, and $\mms\Phi = \matrix{ccc}{\mms\phi_1,\cdot,\mms\phi_N}$ is the matrix of the $N$ mass-normalized eigenvectors defined in terms of the usual orthonormality conditions,
\e{\mms\Phi\tra\mm M \mms\Phi = \mm I\fk\,\mms\Phi\tra\mm K \mms\Phi = \diag\left(\omega_i^2\right)\fp}{figure02}
A modal damping ratio of $D_i=1\%,\,i=1,\ldots,N$ was specified for all modes, if not explicitly stated otherwise.\\

\subsection{Tuning}
The left beam and the right beam have identical material and geometrical properties except for their height $h_i$, as listed below \fref{figure01}.
While the left beam's height $h_1$ remained constant throughout this study, the height $h_2$ of the right beam was varied in order to vary the natural frequencies of its modes, \ie, its component modes in the absence of the joint. Note that all natural frequencies of the beam are proportional to its height. Thus, any ratio between specific natural frequencies of the left and the right beam can be realized by continuously varying $h_2$. In this study, we focused on the $1:3$ internal resonance between the first bending mode of the left beam and the first bending mode of the right beam. To this end, $h_2$ has to be specified to be close to $3h_1$. For convenience, a detuning parameter $\Delta$ is introduced,
\e{h_2\left(\Delta\right) = 3h_1\left(1+\Delta\right)\fp}{detuning}
For $\Delta=0$, the corresponding modes of the left and the right beam are in $1:3$ internal resonance, while for $\Delta> 0$ ($\Delta<0$) the natural frequency ratio between the right and the left beam is larger (smaller) than three.

\subsection{Model reduction\label{sec:model_reduction}}
To reduce the computational effort of the nonlinear dynamic analysis, a conventional modal truncation was utilized. Based on a preliminary convergence study, the six modes $\mms\Phi_{\mathrm{red}}=\matrix{ccc}{\mms\phi_1 &\ldots &\mms\phi_6}$ associated with the six lowest natural frequencies of the system without joint served as reduced basis. In the neighborhood of the $1:3$ internal resonance, these six modes are the leading four modes of the left beam, and the leading two modes of the right one. The associated natural frequencies are $\omega_1=1.0$, $\omega_3=6.27$, $\omega_4=17.5$ and $\omega_6=34.4$ for the left beam, and, assuming $h_2=3h_1$, $\omega_2 = 3.0$ and $\omega_5=18.8$ for the right beam. \eref{eqm} was projected accordingly, resulting in the following reduced set of equations of motion,
\e{\ddot q_i(t) + 2D_i\omega_i\dot q_i(t) + \omega_i^2 q_i(t) + \mms\phi_i\tra\frv\left[\mms\Phi_{\mathrm{red}}\mm q(t)\right] = \pex\mms\phi_i\tra\ewonel\cos\Omega t\fk\quad i=1,\ldots,6\fk}{eqm_modred}
in terms of the modal coordinates $\mm q = \matrix{ccc}{q_1 &\ldots &q_6}\tra$.

\subsection{Influence of the linearized joint behavior}
For sufficiently small magnitudes $\left|\ut\right|\leq\frac{\mu\fnp}{\kt}$ of the joint displacement coordinate $\ut$, the friction contact is only sticking and the joint behaves as a linear spring with stiffness $\kt$. The larger $\kt$ is, the stronger the coupling between the left and the right beam becomes. For $\kt=0$, the two beams are decoupled, and the system modes are isolated to either the left or the right beam, see \fref{figure02}a. For increasing $\kt$, the modes of the linearized system, transform from isolated modes to system modes with increasingly integral character; that is, the vibration energy is more evenly distributed among the two beams, see \fref{figure02}b-c. In the nonlinear regime, a larger $\kt$ also leads to a stronger influence of the nonlinearity on the overall system response. In this study, two different values of $\kt$ were considered: The case of relatively a weak coupling with $\ktweak = 1.25\frac{3EI_1}{l_1^3}$, and the case of a relatively strong coupling with $\ktstrong = 125\cdot\frac{3EI_1}{l_1^3}$. It was found that the mechanism based on nonlinear modal interactions and most of the qualitative results obtained in this work are identical for both weak and strong coupling. Accordingly, most results are presented only for the case of weak coupling with $\ktweak = 1.25\frac{3EI_1}{l_1^3}$. Only in \sref{detuning}, where the resulting friction damping performance is analyzed, results are also presented for the case of strong coupling, owing to the observed important differences between the weakly and strongly coupled cases.
\figw[t!]{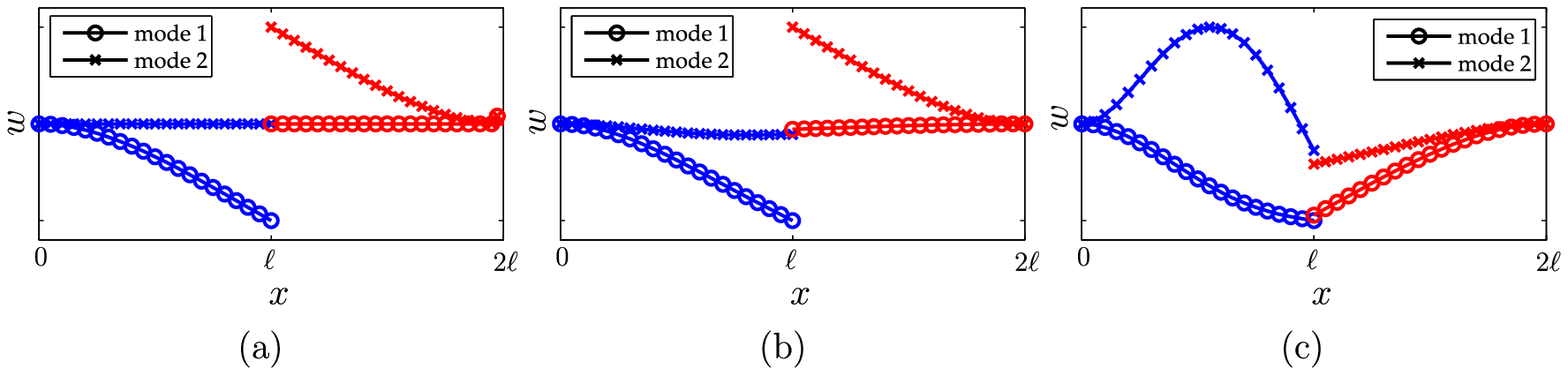}{Vibrational deflection shapes of the first and second linear modes: (a) decoupled with $\kt=0$, $\omega_1 = 1.00$, $\omega_2 = 3.00$, (b) coupled with $\kt = \ktweak := 1.25 \frac{3EI_1}{l_1^3}$, $\omega_1 = 1.45$, $\omega_2 = 3.07$, (c) coupled with $\kt = \ktstrong := 125\cdot\frac{3EI_1}{l_1^3}$, $\omega_1 = 2.56$, $\omega_2 = 4.70$}

\section{Friction-induced nonlinear modal interactions\label{sec:modal_interactions}}
The condition of internal resonance is necessary but not sufficient for the occurrence of an interaction between the corresponding modes. As is demonstrated in this section, the friction nonlinearity can in fact cause this modal interaction and lead to pronounced effects on the forced response. In \ssref{stimvar_frvar}, it is investigated in what dynamic regime this interaction takes place and where it is most prominent. In \ssref{powerflow}, the underlying vibration mechanism is analyzed in terms of the dynamic power flow through the system. Throughout this section, the detuning parameter is set to $\Delta=0$; \ie, the first bending modes of the left and the right beam are exactly in internal resonance condition.

\subsection{Influence of excitation level and limit friction force on the intensity of the nonlinear modal interactions\label{sec:stimvar_frvar}}
\figw[t!]{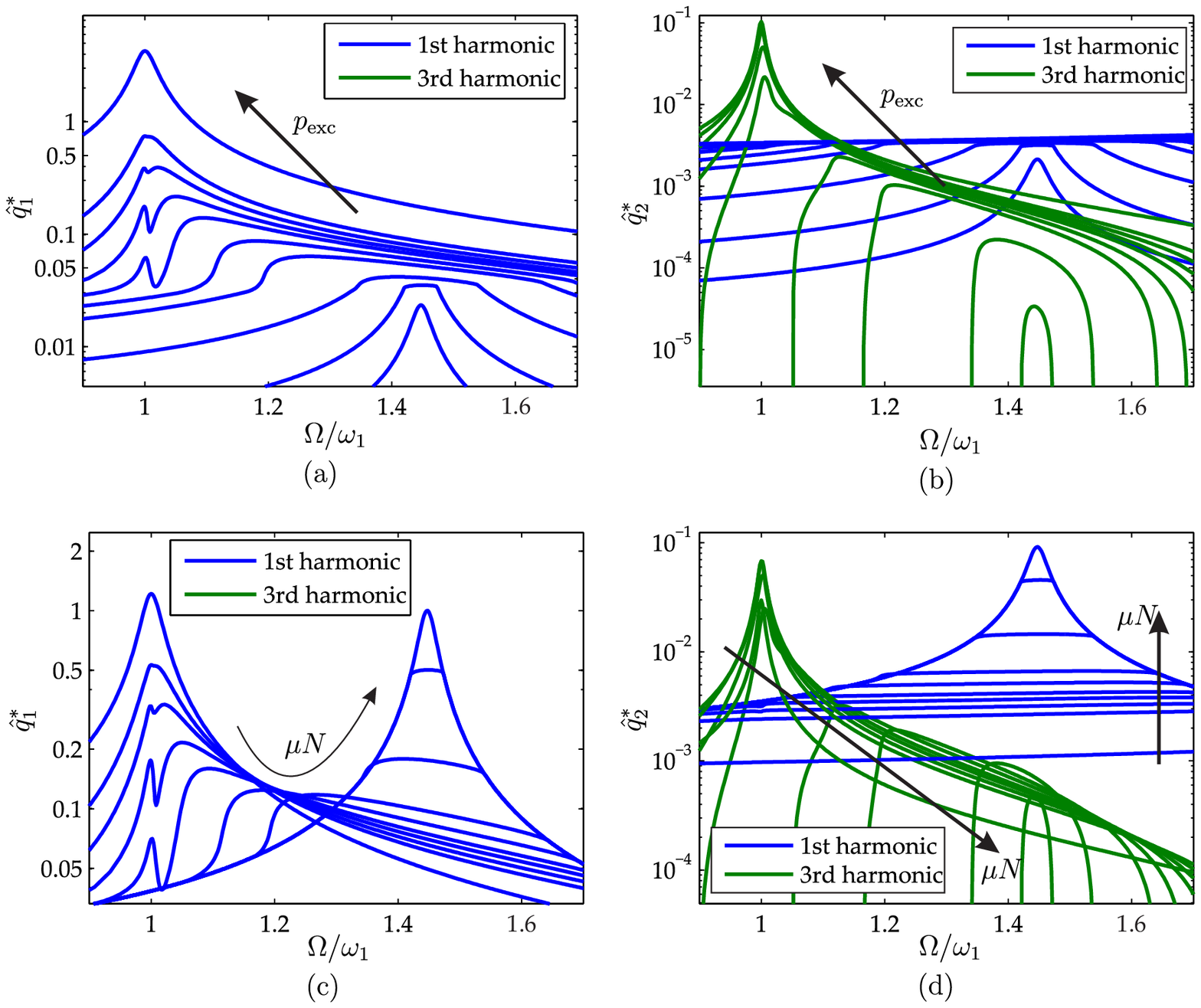}{Normalized steady-state response of the modal amplitudes $\hat q_1^*$, $\hat q_2^*$ of the left and the right beam's first bending mode: (a)-(b) results for left and right beam for variation of the excitation level, $\mu\fnp=1.0$, $ \pex\in\lbrace 0.02,0.07,0.2,0.5,0.7,0.9,1.0,1.2,1.4,3.5 \brace$, (c)-(d) results for left and right beam for variation of the limit friction force, $\mu\fnp \in \lbrace 0.3,0.75,0.9,1.0,1.2,1.5,2.0,4.5,15,45\rbrace$, $\pex=1.0$}
The nonlinear, steady-state forced response is considered in the excitation frequency range in the neighborhood of $\omega_1$, \ie, the natural frequency of the left beam's first bending mode. The vibration response was computed numerically using the well-established harmonic balance method. In the computations, only the first five harmonics were retained in the Fourier expansion of the response of each beam. This number was deemed sufficient in accordance with a preliminary convergence study, where the results were compared with direct numerical time integration. Note that the highest considered natural frequency $\omega_6$ is much larger than the fifth harmonic of the driving frequency $\Omega$, $5\Omega\approx 5.0\ll 34.4 \approx \omega_6$, \cf \ssref{model_reduction}. Hence, the discarded modes are far from the driving frequency and its first five harmonics. The excitation force $\pex$ and limit friction force $\mu\fnp$ were varied over a wide range of values. The results for the modal amplitudes $\hat q_1$, $\hat q_2$ of the left and the right beam's first bending mode are depicted in \fref{figure03} in terms of their first and third harmonic components. The third harmonic of $q_1$ is so small that it lies outside the depicted amplitude range in \fref{figure03}a and c. The results are \textit{normalized} with respect to the resonant case for $\pex=1.0$ and sticking contact conditions. More specifically, the depicted quantities were divided by the magnitude of the first harmonic of $q_1$ for these conditions.\\
Consider first the results depicted in \fref{figure03}a-b for varying excitation level. As stated above, the nonlinear friction joint acts as a spring $\kt$ for small vibrations, so in that limit the system exhibits linear behavior. Since friction damping is absent in this case, the resonance peak is relatively sharp and located at the corresponding natural frequency $\Omega=1.45\omega_1$ of the first system mode. On the other hand, for large vibrations, the friction joint is mostly slipping, but the limited friction force becomes small compared the other forces in the system. Hence, for large vibrations the influence of the friction joint decreases, and the nonlinear system behavior approaches the linear behavior of the system without a joint. Thus, the resonance peak sharpens again and resonance frequency approaches the natural frequency $\Omega=\omega_1$. In between these (quasi-)linear limiting cases, the friction nonlinearity causes a considerable damping effect. As a consequence, the resonance peak broadens significantly, in particular for lower to moderate excitation levels. This behavior is well known for friction-damped systems.\\
At a certain range of the excitation level, considerable effects of nonlinear modal interactions can be ascertained in our results: The third harmonic of the right beam's response assumes a prominent maximum. In this regime, the third harmonic clearly dominates the response of the right beam. In the same range, the peak of the left beam's response transforms into a double-peak. More specifically, the amplitude-frequency curve of the fundamental harmonic component assumes two local maxima with an intermediate local minimum. This phenomenon is a characteristic feature of nonlinear systems in internal resonance.\\
Provided that the elastic Coulomb nonlinearity is the only nonlinearity present in the system, the following scaling property relates the response $\mm u(t)$ of the nonlinear system, the excitation level $\pex$ and the limit friction force $\mu\fnp$ \cite{krac2014b},
\e{\left.\mm u\right|_{\gamma\mu\fnp,\gamma\pex}(t) = \gamma \left.\mm u\right|_{\mu\fnp,\pex}(t)\fk}{scale_invariance}
for any positive real constant $\gamma$. If the response $\mm u(t)$ is known for a set of values for $\mu\fnp$ and $\pex$, \eref{scale_invariance} allows to determine the response for accordingly scaled parameters $\gamma\mu\fnp$ and $\gamma\pex$ without the need for re-computation. As a consequence, the variation of the excitation level $\pex$ has a similar effect as the variation of the limit friction force $\mu\fnp$. Hence, the amplitude-frequency curves depicted in \fref{figure03}c-d are strictly correlated with the ones in \fref{figure03}a-b, and they are merely presented here to more clearly demonstrate the influences of the individual parameters on the steady-state forced responses. 
If the expected excitation level is approximately constant and known a priori, the limit friction force $\mu\fnp$ can be designed to minimize the vibration level. In this case, a reduction by a factor ten can be achieved, compared to the linear case with only sticking joint. The vibration level is less sensitive for values $\mu\fnp$ larger than this nominal optimum, which is usually preferred in order to improve the reliability with respect to uncertainties \cite{krac2014b}.\\
It should be noted that the qualitative dependence of the maximum vibration level on $\pex$ and $\mu\fnp$ is not changed due to the nonlinear modal interactions considered here. Still, the regime of pronounced influence of the modal interactions is close to the optimum design point, and therefore appears relevant to the design process.

\subsection{Analysis of the dynamical mechanism governing the nonlinear modal interactions\label{sec:powerflow}}
\figw[t!]{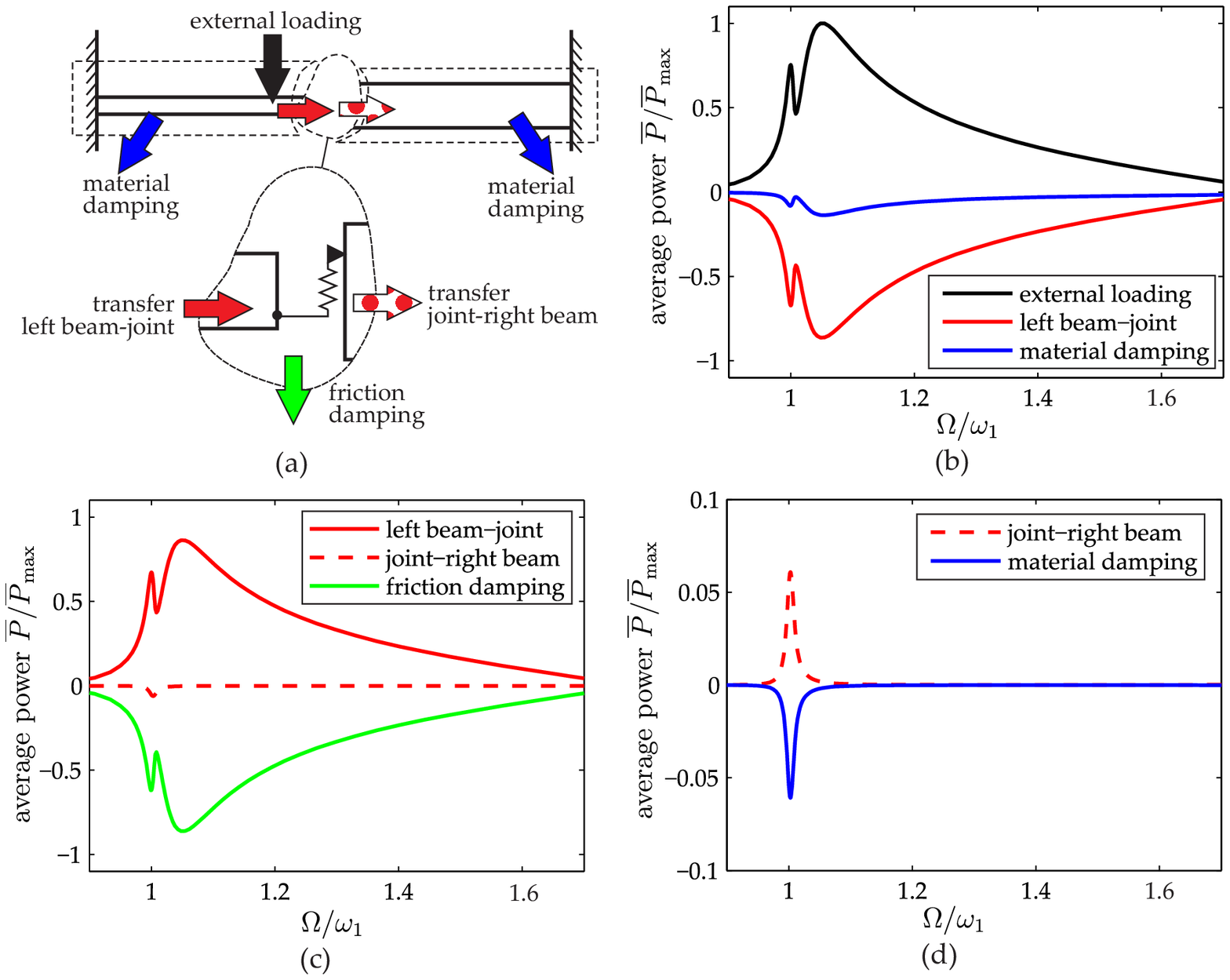}{Energetic analysis of the steady-state vibration response: (a) Schematic illustration of the power flow, (b) power balance in left beam, (c) power balance in joint, (d) power balance in right beam}
In this subsection, we examine the vibration mechanism that leads to emergence of the phenomena associated with nonlinear modal interactions. To this end, we consider the steady-state forced response for a specific excitation level $\pex$ and a specific limit friction force $\mu\fnp$ such that $\mu\fnp/\pex=1.0$. For each system component, \ie, the left beam, the joint and the right beam, we analyze the following averaged mechanical powers: The power input from the external forcing, the power dissipated by material or friction damping and the power transferred to or received from components. These quantities are illustrated in \fref{figure04}. A positive value indicates a power input to, whereas a negative value indicates a power output from the considered component. The orientation of the arrows in the schematic overview \fref{figure04}a takes into account the actual sign of the power.\\
As can be seen in \fref{figure04}b, the left beam is driven by the power of the external forcing. Some of this power is directly dissipated by material damping as a consequence of the structural vibration, but the largest part of the input power is transmitted directly to the frictional joint. In the joint, most of this power is dissipated by dry sliding friction, see \fref{figure04}c. Owing to the nonlinear character of friction, the periodic joint force consists not only of the fundamental frequency component but also of higher frequency components, among which the third harmonic is particularly large. In the neighborhood of the internal resonance, the right beam exhibits a comparatively high dynamic compliance with respect to the third harmonic of the excitation frequency. Hence, power is also transmitted from the joint to the right beam.
Although this power is small compared to the power dissipated in the joint, it is still large enough to maintain the vibration of the right beam. 
The power transmitted to the right beam is, of course, dissipated by the inherent material damping.
\\
From a structural dynamics point of view, two effects of the nonlinear modal interactions are of particular importance: (a) The spatial re-distribution of the vibration energy within the system: While in the absence of modal interactions, the vibration is localized in the left beam, the vibration energy is more evenly distributed between the two beams when nonlinear modal interactions occur. (b) The temporal re-distribution of the vibration energy: While in the absence of modal interactions, the fundamental frequency component clearly dominates the vibration response, considerable energy is transferred to higher frequency components in the case of nonlinear modal interactions.

\section{Resulting friction damping performance and its dependence on frequency detuning\label{sec:detuning}}
In this section, we investigate how a frequency detuning of the system around the considered $1:3$ internal resonance affects the friction damping performance. To this end, the parameter $\Delta$ is varied, which controls the height and therefore the stiffness of the right beam.
From a vibration control perspective, the objective is to maximize the structural reliability and therefore to minimize the mechanical stress in the system. We therefore consider the normal stress $\sigma_{xx}$ in the axial direction in either of the coupled beams. At a given spatial location within the structure, the stress varies with time $t$. The steady-state, periodic forced response generally exhibits multi-harmonic character. We thus consider its root-mean-square (RMS) value of the stress,
\e{\sigrms\left(x_i,y\right) = \sqrt{ \frac{1}{T} \int\limits_{0}^{T} \sigma_{xx}^2\left(x_i,y,t\right) \dd t}\,\,i=1,2\fp}{sigrms}
where $x_i$ is the local axial coordinate for each beam, $y$ is the transverse coordinate, and $T=\frac{2\pi}{\Omega}$ is the fundamental period of oscillation.\\
In the absence of normal loading, $\sigma_{xx}(x_i,y,t)=Ew^{\prime\prime}(x_i,t)y$. The critical transversal location is thus always the top or the bottom of the rectangular cross-sectional area, $y=\pm h_i/2$, of each beam. The critical stress within each beam is thus
\e{\sigcriti = \max\limits_{0\leq x_i\leq\ell} \sigrms\left(x_i,\frac{h_i}{2}\right)\fk\,\, i=1,2\fp}{sigcritonetwo}
The critical cross-sectional area is typically located at the clamped end, \ie, at $x_1=0$ and $x_2=\ell$ for the right and the left beam, respectively. Hence, the critical stress within the system is the largest critical component stress,
\e{\sigcrit = \max\left(\sigcritone, \sigcrittwo\right)\fp}{sigcrit}
Owing to the comparatively small vibration level of the right beam, $\sigcrit$ is \textit{always attained in the left beam}, that is $\sigcrit=\sigcritone$ in all cases presented in this study. In the case of steady-state vibrations in the presence of harmonic external forcing, the critical stress $\sigcrit$ is a function of the excitation frequency $\Omega$, and the maximum $\sigcrit$ within the considered frequency range is denoted by $\performance$,
\e{\performance = \max\limits_{\Omega} \sigcrit\left(\Omega\right)\fp}{objective_function}
This quantity is used to assess the effect of frictional damping on the fatigue of the coupled structure, and is thus defined as the objective function in the subsequent numerical studies.\\
For sufficiently large $\left|\Delta\right|$, the nonlinear modal interactions with the right beam are suppressed, and the response level approaches an (almost) constant value. This \textit{detuned} case serves as a reference. Stress values are accordingly normalized by the maximum (resonant) stress $\performanceref$ in this case. As it turns out, some quantitative and qualitative differences between weak and strong coupling can be ascertained. The case of weak coupling is considered first.

\subsection{Weak coupling\label{sec:weak}}
\figw[t!]{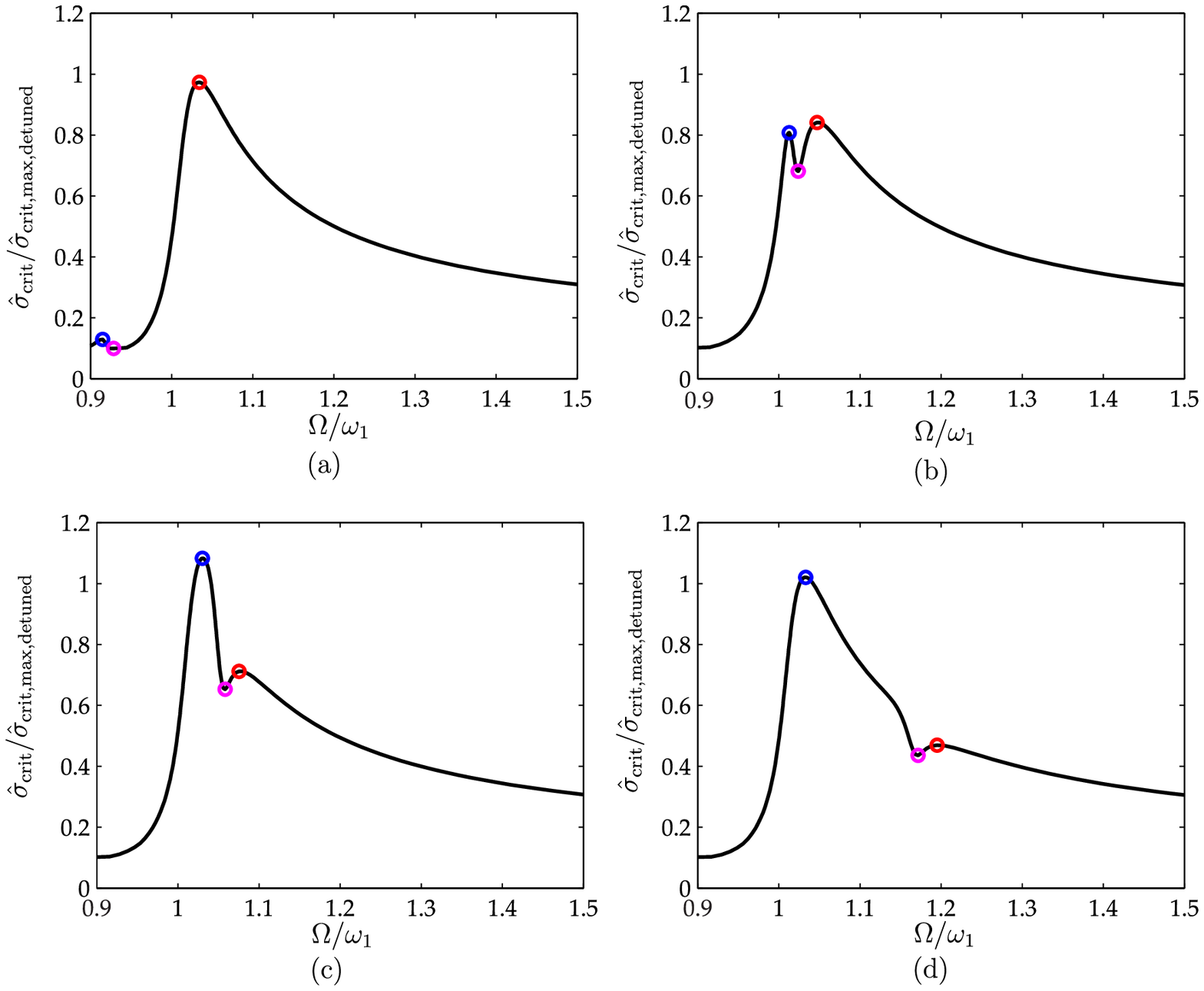}{Steady-state response of the normalized critical stress for different detuning values: (a) $\Delta=-0.09\omega_1$, (b) $\Delta=0.017\omega_1=:\Deltaopt$, (c) $\Delta=0.05\omega_1=:\Deltaworst$, (d) $\Delta=0.16\omega_1$}
%
The detuning parameter $\Delta$ was varied in the neighborhood of $\Delta=0$. For several values of the detuning, the critical stress $\sigcrit$ is shown as a function of the excitation frequency in \fref{figure05}. In all four cases, the critical stress response exhibits two local maxima and an intermediate minimum. This is similar to the results depicted in \fref{figure03}a and c for the case $\Delta=0$. The location of the local minimum and the two local maxima depends on $\Delta$. As $\Delta$ is increased, the location of the minimum is shifted to larger frequencies. The domain of influence of this phenomenon is limited to a comparatively small frequency window. For larger $\left|\Delta\right|$, as in \frefs{figure05}a and d, the dominant peak occurs around $\Omega=1.03\omega_1$, and approaches the level of the detuned case. The overall effect of the nonlinear modal interactions is most pronounced when the local minimum is located at $\Omega\approx 1.03\omega_1$, \ie, for $\Delta\approx 0.03$.
\figw[t!]{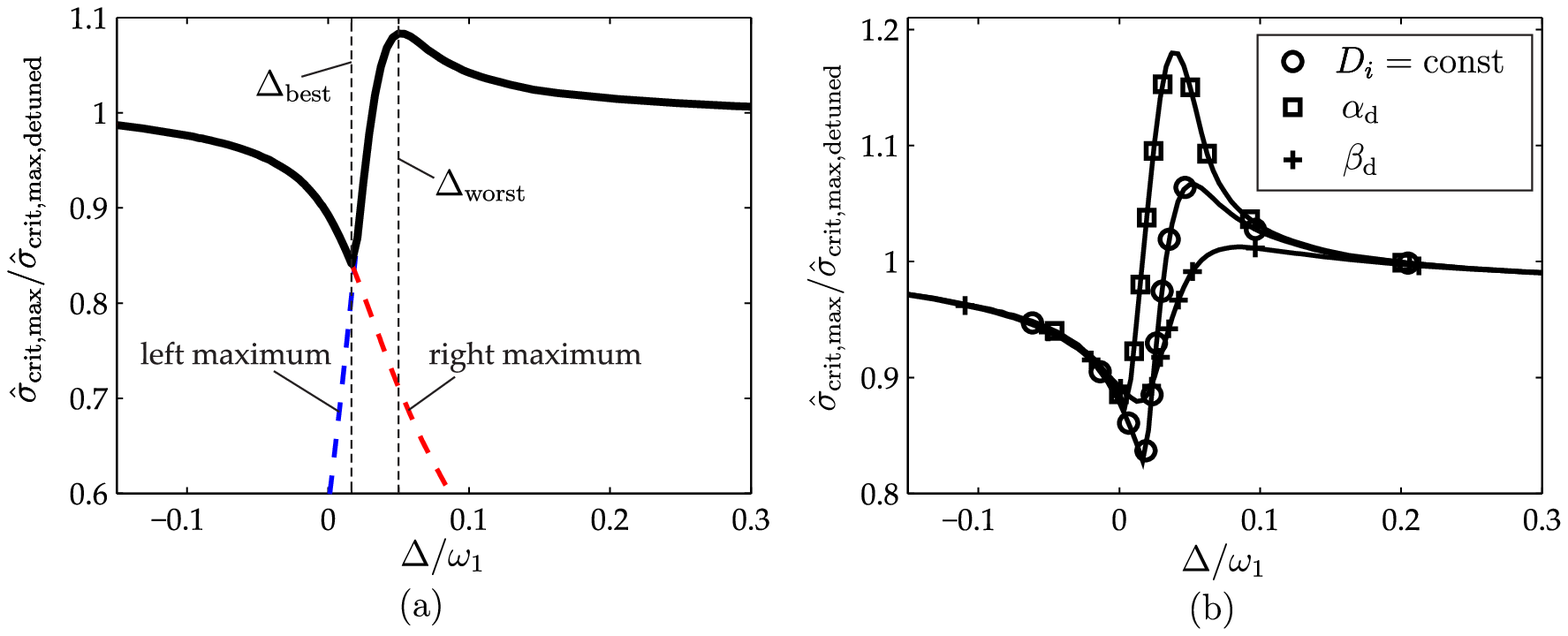}{Dependence of the maximum critical stress on detuning: (a) Global maximum of the steady-state critical stress, composed of the two local maxima as indicated in \fref{figure05}, (b) influence of the type of the linear viscous damping}
\\
Not only the location of the secondary peaks varies, but also their height depends on $\Delta$. For smaller $\Delta$, the right secondary peak is larger than the left one. For a certain value of $\Delta=0.017\omega_1$ both peaks are equal. For larger $\Delta$, the left peak is larger than the right one. This can be seen well in \fref{figure06}, which shows the relationship $\performance(\Delta)$ between the maximum stress level $\performance$ within the considered frequency range and the detuning parameter $\Delta$. For a certain $\Delta$, the stress level is minimized. The associated $\Delta=\Deltaopt$ is referred to as optimum tuning. This corresponds to the point where the secondary peaks have equal stress level, see \fref{figure05}b. Since the critical frequency point switches at this point from the right to the left critical frequency, the characteristic has a kink at $\Delta=\Deltaopt$. The height of the right maximum shrinks monotonically for increasing $\Delta$. In contrast, the left maximum grows quickly with $\Delta$ up to a certain point $\Deltaworst$, and then shrinks slightly again. Remarkably the stress level at $\Deltaworst$ is larger than in the detuned case. In other words, \textit{the presence of an internal resonance has detrimental effects on the friction damping performance}. An explanation for this possibly unexpected phenomenon is given in \ssref{detrimental}.\\
The nonlinear modal interactions rely on the vibration energy transfer from low to high frequencies. The energy dissipation provided by material damping, which is assumed to be of viscous linear type in this work, depends on the vibration frequency: The higher the vibration frequency, the larger is the amount of energy dissipated within a given time span. That is why viscous damping is more efficient for higher vibration frequencies. While the fundamental frequency component is largely confined in the left beam's first mode, the higher frequency component is largely confined in the right beam's first mode. The damping ratios associated with these two modes can have a significant influence on the overall dynamic behavior of the system. 
A modal damping ratio $D_i=1\%$ for all modes was assumed, as defined in \eref{modal_damping}. In order to demonstrate that the qualitative behavior shown in \fref{figure06}a is invariant under the variation of the assumed damping model, two different cases are considered: Mass-proportional damping with $\mm C=\alphad\mm M$, and stiffness-proportional damping with $\mm C=\betad\mm K$. Generally, mass-proportional damping results in lower damping of higher frequency components, while the opposite holds for stiffness-proportional damping. The coefficients $\alphad$, $\betad$ were specified to maintain equal modal damping ratio of the first mode $D_1=1\%$ regardless of the damping model. Hence, the damping ratio $D_2$ of the interacting higher-frequency mode depends on the damping model: For $\Delta=0$, one has $D_2=1/3$ for mass-proportional damping, $D_2=D_1=1\%$ for equal modal modal damping, and $D_2=3\%$ for stiffness-proportional damping. The results are depicted in \fref{figure06}b. It can be clearly seen that the qualitative dependence on $\Delta$ does not change. However, quantitative differences exist that are caused by two different effects: (a) The better the higher frequency component is damped, the lower is the worst stress level (that is the largest value of $\performance$). (b) A stronger damping of the higher frequency component also reduces the response of the higher frequency mode, and may even suppress the nonlinear modal interactions. As a consequence, the stress-detuning curve flattens out for larger damping of the higher frequency components.

\subsection{Strong coupling\label{sec:strong}}
We now consider the case of a much larger joint stiffness $\ktstrong = 125\cdot\frac{3EI_1}{l_1^3}=100\ktweak$, which is $100$ times larger than the coupling stiffness considered in the previous sections. A larger joint stiffness results in a stronger coupling between the two beams, and, hence, a more pronounced influence of the nonlinearity. As in the case of $\ktweak$, two quasi-linear limit cases exist for small and for large excitation levels, respectively, as can be ascertained from \fref{figure03}. The shift between the natural frequencies for sticking and free contact conditions is much larger for $\ktstrong$, and the dynamics in the intermediate regime between the quasi-linear limit cases becomes much more complicated. The first $21$ Fourier terms were required in the harmonic balance method, compared to the five terms in the previous case, in order to achieve satisfactory agreement with direct numerical time integration. We would like to emphasize that the \myquote{noisiness} in the intermediate regime is not due to the harmonic truncation but was also predicted by the numerical time integration. Instead, this behavior is considered intrinsic to the system. Due to the strong coupling, the non-smooth character of the stick-slip transitions is no longer filtered out by the structure, but propagates to the global vibration behavior. For moderate and high excitation levels, the system again exhibits the most pronounced response in the neighborhood of $\Omega\approx\omega_1$. Here, the dynamic behavior is comparatively regular and dominated by the first component mode of the left and the right beam, associated with the modal coordinates $q_1$ and $q_2$, respectively. The vibration energy is mainly confined to the first and third harmonic, as in the previous case, while the response of the higher harmonics (not illustrated) is comparatively small. In the regime of higher excitation levels, it is thus still reasonable to investigate the nonlinear interaction between the first two modes.
\figw[t!]{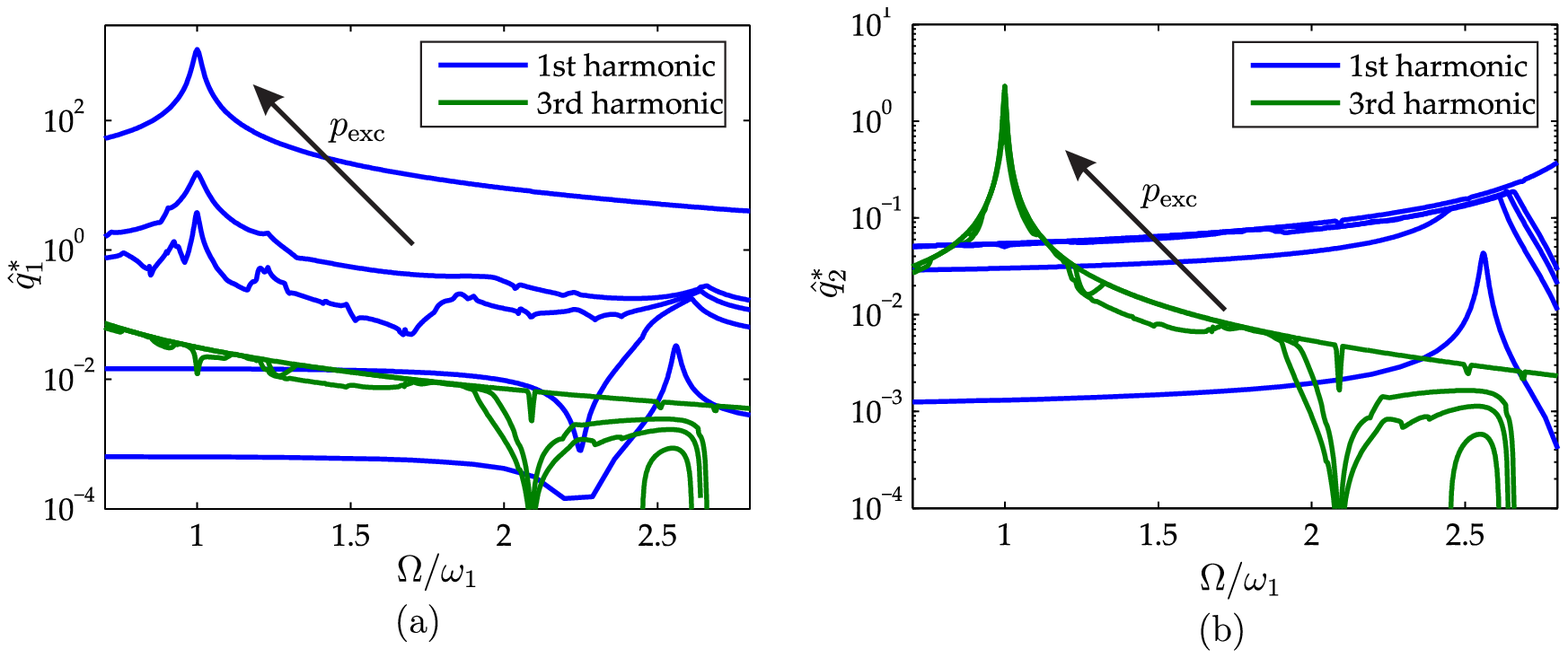}{Steady-state response of the beams' tip displacement amplitudes for variation of the excitation level and strong coupling, $\Delta=0$: (a) Left beam, (b) right beam}
\figs[th]{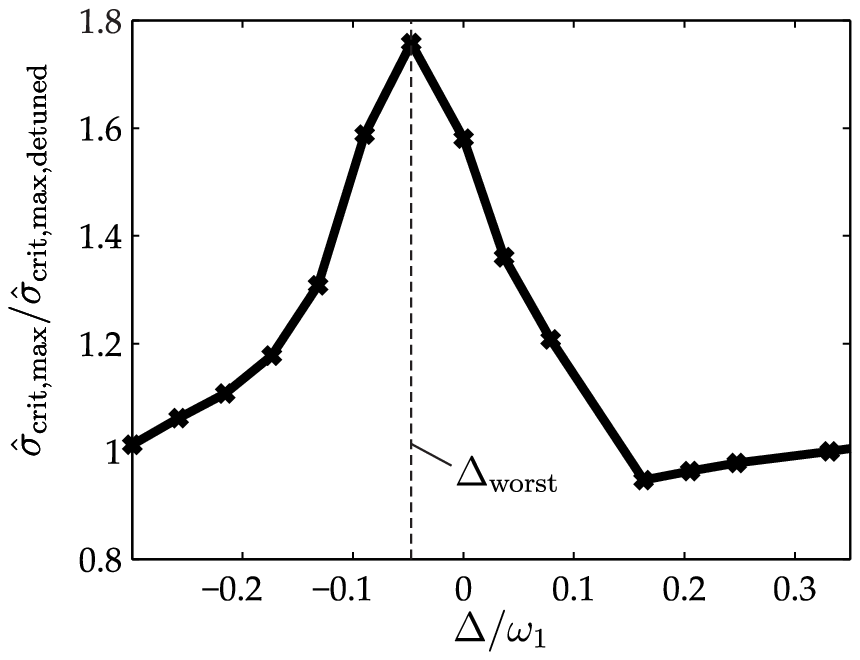}{Dependence of the maximum critical stress on detuning for strong coupling}{1.0}
\\
We now consider the third-largest excitation level depicted in \fref{figure07}. The detuning parameter $\Delta$ was varied, and the steady-state forced response was computed in the frequency range $0.5\omega_1\leq\Omega\leq2.8\omega_1$. The resulting maximum stress level $\performance$ is depicted in \fref{figure08}. The qualitative dependence of $\performance$ on $\Delta$ differs considerably from the results obtained for the case of weak coupling shown in \fref{figure06}. There is a qualitative difference, since the nonlinear modal interactions are only detrimental regarding the maximum stress level attained in the system. Hence, in the case of strong coupling, the harmonically forced system cannot be tuned to perform better due to the $1:3$ internal resonance. Moreover, there are also quantitative changes, as the nonlinear modal interactions cause a variation of the stress level $\performance$ of $\approx\pm 15-20\%$ for $\ktweak$, whereas the stress level is more than doubled for $\ktstrong$. The quantitative discrepancy is obviously caused by the increased relevance of the nonlinearity on the overall dynamics, owing to the larger joint stiffness $\kt$. The qualitative difference is less obvious. This is addressed in the following subsection.

\subsection{On the detrimental effects of the nonlinear modal interactions\label{sec:detrimental}}
As was ascertained both for the cases of weak and strong coupling, the nonlinear modal interactions caused by the $1:3$ internal resonance between the two coupled beams can have detrimental effects on the friction damping performance. This, probably, is an unexpected outcome. Apparently, the presence of the higher frequency component (mainly in the right beam) in the vibration response reduces the effective damping of the fundamental frequency component (mainly concerning the left beam). Note that damping of the fundamental frequency component is most important, since the critical response level was always reached in the left beam, whose steady-state response is dominated by the fundamental frequency component. In what follows, the effective damping provided by the elastic Coulomb friction element is investigated in more detail, in order to confirm the aforementioned assertions.\\
First, we define a measure for the effective damping of the fundamental frequency component. To this end, consider a periodic, two-harmonic joint displacement $\ut(t)$,
\e{\ut(t) = \real{ a_1\ee^{\ii\Omega t} + a_3\ee^{\ii\delta_{31}}\ee^{3\ii\Omega t} }\fp}{joint_displacement}
Herein, $a_1$ and $a_3$ are the amplitudes of the first and third harmonic, and $\delta_{31}$ is the possible phase lag between the harmonic components. 
The following effective damping coefficient is defined,
\e{\deffone = \frac{\left(\ii\Omega a_1\right)^*~\frac{1}{2\pi}\int\limits_{(2\pi)}^{}\fr\left[\ut\left(\Omega t\right)\right]\ee^{-\ii\Omega t}\dd\Omega t}{\Omega^2 a_1^2}\fp}{deffone}
Herein, $^*$ denotes the complex conjugate, and $\fr\left[\ut\left(\Omega t\right)\right]$ is the nonlinear force characteristic of the elastic Coulomb element. It can be verified that if $\fr$ was replaced by a linear viscous damper force with coefficient $d$, \eref{deffone} would yield the expected value $\deffone=d$.\\
The elastic Coulomb model has two parameters, namely $\mu\fnp$ and $\kt$. The model possesses scaling properties with respect to these parameters as follows,
\ea{\left.\fr\left[\ut\right]\right|_{\gamma\mu\fnp} = \gamma\left.\fr\left[\frac{1}{\gamma}\ut\right]\right|_{\mu\fnp}\fk\label{eq:scalinga}\\
\left.\fr\left[\ut\right]\right|_{\gamma\kt} = \left.\fr\left[\gamma\ut\right]\right|_{\kt}\fk\label{eq:scalingb}}{scaling}
for arbitrary $\gamma\in\mathbb R_+$. In order to obtain a complete picture of the behavior of $\deffone$, it is thus sufficient to keep $\kt$ and $\mu\fnp$ fixed and to determine the dependence of this measure only on $a_1$, $a_3$ and $\delta_{31}$. From the resulting functional relationship, $\deffone$ can then be reproduced for any given value for $\kt$ and $\mu\fnp$ by utilizing \erefs{scalinga}-\erefo{scalingb}.
\figw[t!]{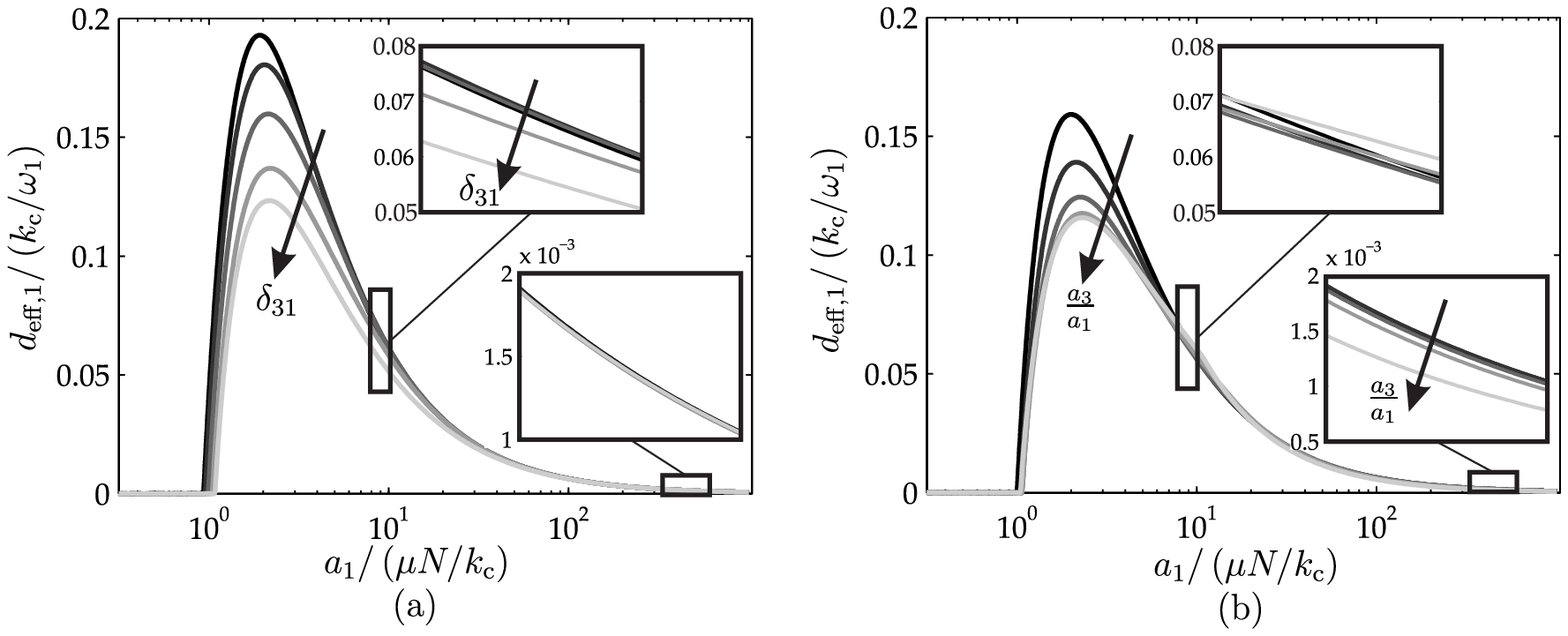}{Effective damping of the elastic Coulomb element provided to the first harmonic, versus its amplitude $a_1$: (a) variation of the phase lag between the harmonic components $0\leq\delta_{31}\leq\pi$, for fixed $\frac{a_3}{a_1}=0.075$; (b) variation of the amplitude ratio between the harmonic components $0 \leq\frac{a_3}{a_1}\leq 0.2$, for fixed $\delta_{31}=0.85\pi$; the left and right zooms correspond to the relevant amplitude ranges for weak and strong coupling, respectively}
\\
The dependence of the effective damping measure $\deffone$ on the parameters $a_1$, $\frac{a_3}{a_1}$ and $\delta_{31}$ was studied in detail. The investigation was limited to those parameter ranges that are relevant for the results reported in this work. To this end, the forced response depicted in \fref{figure06} and \fref{figure08} was considered and the corresponding values for $a_1$, $\frac{a_3}{a_1}$ and $\delta_{31}$ were determined. The results of this analysis are not presented here for the sake of brevity, and the measure $\deffone$ was computed within the relevant parameter ranges using \eref{deffone}. The results are illustrated in \fref{figure09}a-b.\\
The evolution versus $a_1$ qualitatively resembles previously reported results obtained in the absence of nonlinear modal interactions, \eg, \cite{whit1996,laxa2009,Krack.2015a}. We therefore limit the discussion to the influence of the higher harmonic component. The left and right zooms in each sub-figure in \fref{figure09} correspond to the relevant ranges of the amplitude $a_1$ for weak and strong coupling, respectively. Apparently, different properties of the vibration response cause the detrimental reduction of $\deffone$ in the cases of weak and strong coupling.
\begin{itemize}
\item In the case of weak coupling, the phase lag $\delta_{31}$ of the third harmonic plays the central role, almost independent of its magnitude $a_3$. Under the condition of in-phase (anti-phase) joint motion, in a multi-harmonic sense, the elastic Coulomb element is most (least) effective.
\item In contrast, the third harmonic amplitude of the input joint displacement, $a_3$, is crucial in the case of strong coupling, regardless of its phase. The larger $a_3$, the smaller the effective damping of the elastic Coulomb element.
\end{itemize}
It is well-known that the amplitude ratio $\frac{a_3}{a_1}$ commonly undergoes considerable variation in the forced response in the neighborhood of an internal resonance. Recent investigations by the authors suggest that the same is true for the so-called motion complexity, \ie, the phase difference between the coordinates of the nonlinearly interacting modes \cite{Krack.2015b,Krack.2015c}.

\section{Conclusions\label{sec:conclusions}}
The concept of friction damping relies on the inherently nonlinear effects of dry friction. The nonlinear forces can trigger nonlinear interactions among (nearly) internally resonant modes, in spite of the considerable damping that is present in the system. The influence on the steady-state forced vibrations is particularly pronounced in the neighborhood of the optimum design point, and therefore appears relevant for design of friction-damped systems. Depending on the system properties, it is indeed possible to take advantage of the energy redistribution caused by the modal interactions, and to optimally tune the system to achieve a minimum response level. In the numerical study, a reduction of the maximum response level on the order of $10\%$ was achieved, which is deemed relatively small. More importantly, the vibration behavior is much more sensitive, and thus susceptible to inherent uncertainties, as compared to the detuned case in the absence of modal interactions. Moreover, parameter regimes exist, for which the nonlinear modal interactions are detrimental. Apparently, the presence of higher frequency components in the relative joint displacement can diminish the effective damping provided by the friction joint. In the last example, the response level was almost doubled. It is thus concluded that the condition of internal resonance should be avoided in the design of friction-damped systems.




\end{document}